# Evolutionary genetic optimization of the injector beam dynamics for the ERL test facility at IHEP

JIAO Yi (焦毅)

*Institute of High Energy Physics, CAS, Beijing 100049, P.R. China*

**Abstract**: The energy recovery linac test facility (ERL-TF), a compact ERL-FEL (free electron laser) two-purpose machine, was proposed at the Institute of High Energy Physics, Beijing. As one important component of the ERL-TF, the photo-injector started with a photocathode direct-current gun was designed and preliminarily optimized. In this paper an evolutionary genetic method, non-dominated sorting genetic algorithm II, is applied to optimize the injector beam dynamics, especially in the high-charge operation mode. Study shows that using an incident laser with rms transverse size of 1~1.2 mm, the normalized emittance of the electron beam can be kept below 1 mm.mrad at the end of the injector. This work, together with the previous optimization for the low-charge operation mode by using the iterative scan method, provides guidance and confidence for future constructing and commissioning of the ERL-TF injector.

## 1 Introduction

The energy recovery linac (ERL) and free electron laser (FEL) are considered to be candidates of the fourth generation light sources, and have received much attention worldwide. Since both of them are based on linac technologies, it is possible to combine FEL into an ERL facility, resulting in a compact two-purpose light source. A test facility, named energy recovery linac test facility (ERL-TF), was proposed at the Institute of High Energy Physics, Beijing, to verify this principle [1]. Physical design of the ERL-TF started a few years ago and is well in progress [2-5]. The layout and main parameters of the facility are presented in Fig. 1 and Table 1, respectively. Among the components of the test facility, one extremely important device dominating the machine performance is the photo-injector. The injector, including a 500-kV photocathode direct-current (DC) gun equipped with a GaAs cathode, a 1.3 GHz normal conducting RF buncher, two solenoids, and two 2-cell superconducting RF cavities, was designed for the ERL-TF [2], with the layout shown in Fig. 2. Using an incident laser with rms transverse size $\sigma_{laser}$ of 1.2 mm, the designed injector in high-charge operation mode (bunch charge 77 pC, rep. rate 130 MHz) was simulated with the ASTRA program, and finally an electron beam, with kinetic energy $E_k$ of 5 MeV, normalized emittance $\varepsilon_{n,x(y)}$ of 1.49 mm.mrad, rms bunch length $\sigma_z$ of 0.67 mm and rms energy spread $\sigma_\delta$ of 0.72%, was achieved at the end of the injector.

Recently continuous efforts have been made to further optimize the injector beam dynamics based on the simulations with the Impact-T program [6], a fully 3D program to track relativistic particles taking into account space charge force and short-range longitudinal and transverse wake-fields. The beam dynamics of the injector in the low-charge operation mode (bunch charge 7.7 pC, rep. rate 1.3 GHz) was optimized with the iterative scan method. The beam parameters after optimization were $E_k$ = 5 MeV, $\varepsilon_{n,x(y)}$ = 0.4 mm.mrad, $\sigma_z$ = 0.74 mm and $\sigma_\delta$ = 0.33% by using an incident laser with $\sigma_{laser}$ of 0.5 mm. In addition, it was found that the optimized result had rather high tolerance to the parameter fluctuation, magnetic and alignment errors (For more detail, see Ref. [5]).

However, when applying the iterative scan method to the optimization for the high-charge operation mode, it turns to be difficult to achieve a promising beam quality in a moderate period of time, due to higher electron density and stronger space charge effect. Note that the injector beam dynamics optimization is a highly constrained multi-objective optimization problem, and one can use evolutionary genetic algorithm to find globally optimal solutions for such a problem (see, e.g., [7-10]). Therefore in this paper a genetic algorithm, non-dominated sorting genetic algorithm II (NSGA-II [11]), is applied to optimize the injector beam dynamics in both the low-charge and high-charge operation modes. In this study totally twelve parameters are varied and three objectives, $E_k$, $\varepsilon_{n,x(y)}$ and $\sigma_z$, are optimized. The goal is to obtain electron beam with $E_k$ of 5 MeV, $\sigma_z$ of 2 - 4 ps (i.e., 0.6 - 1.2 mm), and $\varepsilon_{n,x(y)}$ as low as possible at the end of the injector. For the low-charge operation mode with $\sigma_{laser}$ of 0.5 mm, the algorithm has a fast convergence within evolution over 50 generations; moreover, it shows that the result obtained with the iterative scan method is very close to the so-called Pareto optimal front of the objectives. However, for the high-charge operation mode the convergent rate of the algorithm is relatively slow. This explains why it is difficult to find a satisfying result for the high-charge operation mode by iteratively scanning the parameters. As a result, the random seeds are evolved over more generations. In addition, the dependency between the available minimum $\varepsilon_{n,x(y)}$ and $\sigma_{laser}$ is investigated for the high-charge operation mode. It is found that using a driven laser with $\sigma_{laser}$ of 1 ~ 1.2 mm helps to achieve an electron beam with $\varepsilon_{n,x(y)}$ below 1 mm.mrad at the end of the injector.

In the following, the NSGA-II algorithm will be described in Sec. 2, and the application of this algorithm in the injector beam dynamics optimization is shown in Sec. 3. Conclusions are given in Sec. 4.

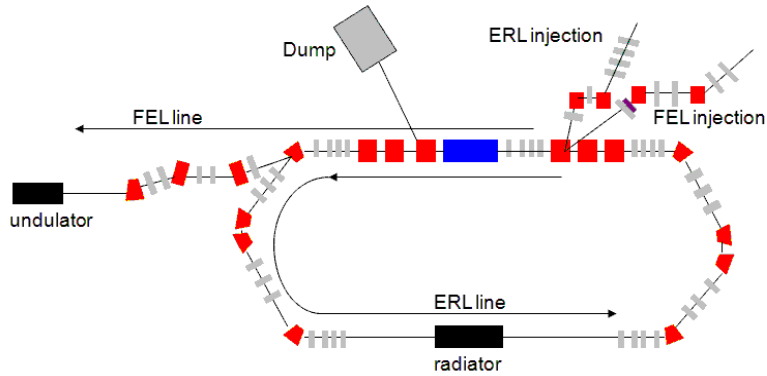

Fig. 1. Layout of the ERL test facility at IHEP.

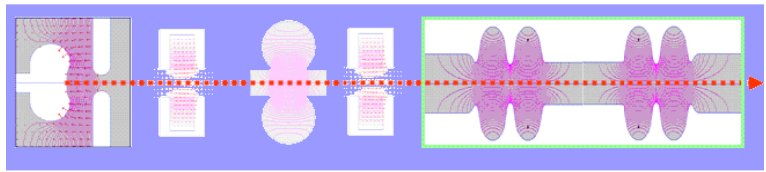

Fig. 2. Layout of the ERL-TF injector, consisting of, from left to right, DC-gun, the first solenoid, RF buncher, the second solenoid, and two 2-cell RF cavities.

Table 1. Main parameters of the ERL-TF at IHEP

| Parameter | Value |
|---|---|
| Beam energy (MeV) | 35 |
| Beam current (mA) | 10 |
| Bunch charge (pC) | 77 (or 7.7) |
| Normalized emittance (mm.mrad) | <2.0 (or < 1.0) |
| Rms bunch length (ps) | 2.0-4.0 |
| Rms energy spread (%) | 0.2-1.0 |
| Bunch frequency (MHz) | 130 (or 1300) |
| RF frequency (MHz) | 1300 |

## 2 NSGA-II algorithm and its concrete implementation

In a multi-objective optimization problem, usually a number of parameters with specific variable ranges are needed to be determined, the objectives may be in conflict, and in the objective space the solutions may be discontinuous. Therefore it is always not possible to find a single solution that optimizes all the objectives simultaneously. To dealing with this challenge, evolutionary genetic methods are usually used to find the so-called Pareto optimal front that represents the set of solutions showing all the possible tradeoffs between the different objectives. The NSGA-II algorithm is such a genetic method. It was demonstrated that the Pareto optimal front obtained by this algorithm converges to the real optimal front for some test problems [11].

The NSGA-II algorithm mimics the nature select: At first, a random population with $N$ individuals is generated and evaluated. Then the parents are chosen from the population according to the rank and crowding distance, where the rank represents the non-dominance of one individual by others and the crowding distance gives a measure of how close an individual is to its neighbors. An individual with less rank or greater crowding distance than others has priority to be selected. The selected parents generate offsprings from crossover and mutation. The objective functions are evaluated on current offsprings, the offsprings together with parents are sorted again based on their ranks and crowding distances, and only the best $N$ individuals are selected. This procedure repeats generation by generation, until reaching a generation with the desired convergence to the Pareto optimal set. More details of the NSGA-II algorithm can be found in Ref. [11].

In this study the NSGA-II algorithm program runs in Matlab on a single PC with multi-threading processors, which makes it available to start several runs of Impact-T simulations simultaneously. The population size of each generation is chosen to $N = 350$, as a compromise between the comprehensiveness of the solutions and the computing time that increases with the population size. It takes about three hours to finish the simulations for one generation. Totally twelve parameters, including the positions, strengths, RF phases (if have) of the injector elements, are varied to investigate the optimal tradeoffs between the different beam parameters at the end of the injector. Three objectives are set, $\varepsilon_{n,x(y)}$, $|E_k -5 \text{ MeV}|$, and $|\sigma_z -0.85 \text{ mm}|$ with the goal to obtain electron beam with $E_k$ close to 5 MeV, $\sigma_z$ close to 0.85 mm, and $\varepsilon_{n,x(y)}$ as low as possible. To avoid loss of possible optimal parameter settings, the variable range of each parameter is set to as large as possible, e.g., −180 to 179 degree for the RF phase. For each parameter setting, the input file for the Impact-T is generated automatically, and then is put into simulation to evaluate the objectives.

The electron beam is created at the GaAs cathode driven by a 532-nm laser, with round cross section and longitudinal beer-can profile. It is assumed that the initial electron beam has the same profile as the laser in transverse planes and in *z* dimension (with flat top of 20 ps, rise and fall time of 2 ps), while has a uniform kinetic energy distribution between 0 and 0.4 eV, with an average of 0.2 eV. The initial normalized emittance or the thermal emittance is given by

$$\varepsilon_{n,x(y)} = \sigma_{x(y)} \sqrt{\frac{k_B T_\perp}{m_e c^2}}, \quad (1)$$

where $\sigma_{x(y)} = \sigma_{laser}$, $m_e c^2$ is the electron rest energy, and $k_B T_\perp$ is the transverse beam thermal energy that depends mainly on the incident laser wavelength [12],

$$k_B T_\perp (meV) = 309.2 - 0.3617 \lambda (nm). \quad (2)$$

In our case $\lambda = 532$ nm and $k_B T_\perp = 116.8$ meV.

**3 Injector Beam Dynamics Optimization with NSGA-II**

In the optimization for the low-charge operation mode, only the case with $\sigma_{laser}$ of 0.5 mm is considered. The population with 350 random seeds evolves over 100 generations and converges to the Pareto front. For the solutions in each generation, we count the minimum emittances under three conditions: (1) without any limitation on $E_k$ and $\sigma_z$; (2) with $|\sigma_z - 0.85| < 0.4$ mm; (3) with $|\sigma_z - 0.85| < 0.4$ mm and $|E_k - 5| < 0.1$ MeV. Fig. 3 shows the variation of the minimum emittances with the generation index. The minimum emittance under condition (3) becomes very close to that under condition (1) after 50 generations, with the difference less than 0.03 mm.mard. The results of the 100th generation in the objective space are shown in Fig. 4. One can see that the solution space is not continuous. This discontinuity makes it impossible to use traditional linear scan methods to get the whole Pareto front. Nevertheless the results in the region labeled 'B' in Fig. 4 all have $\sigma_z$ larger than 4 ps, thus they will be not considered as candidates of the satisfying results in this study. We show only the results satisfying condition (3) in the last 10 generations as well as the result obtained by iterative scanning in Fig. 5. It shows that the optimized result with the iterative scan method is close to the Pareto front.

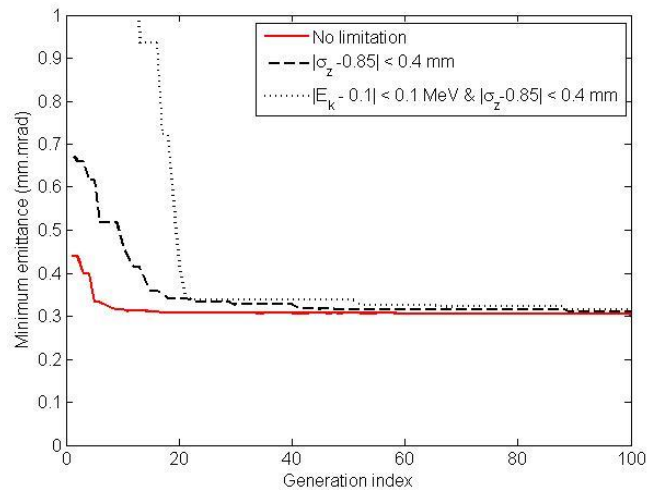

Fig. 3. Variation of the minimum emittances under three different conditions for the low-charge operation mode with $\sigma_{laser}$ of 0.5 mm.

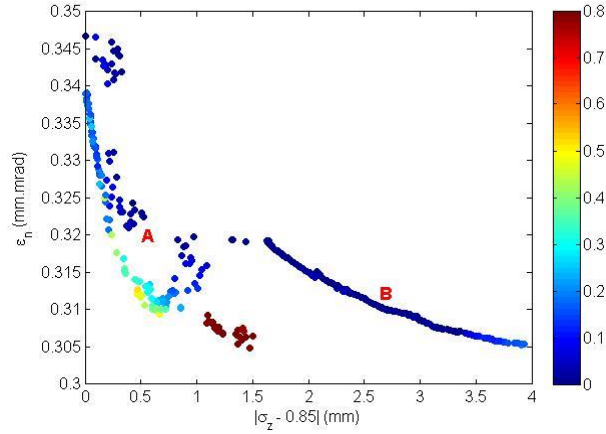

Fig. 4. (color online) Results of the 100th generation in objective space for the low-charge operation mode with $\sigma_{laser}$ of 0.5 mm.

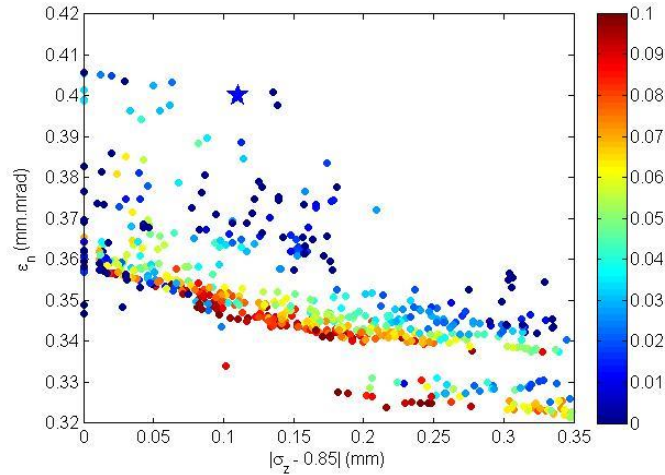

Fig. 5. (color online) Results satisfying conditions $|\sigma_z -0.85| < 0.4$ mm and $|E_k -5| < 0.1$ MeV in the last 10 generations for the low-charge operation mode with $\sigma_{laser}$ of 0.5 mm. The result obtained previously with the iterative scan method is also plotted as a star.

In the optimization for the high-charge operation mode, the population converges relatively slowly to the Pareto front. Taking the case with $\sigma_{laser}$ of 0.75 mm as example, the evolution of the minimum emittances under the above three conditions is shown in Fig. 6. The difference between the minimum emittance under condition (1) and that under condition (2) and (3) is still large even with evolution over 200 generations, ~ 0.5 mm.mrad. This large difference can be understood from the view of the results of the 200th generation in the objective space (Fig. 7). There are three distinct regions in the figure labeled 'A', 'B' and 'C'. The solutions in the region 'B' and 'C' predict smaller emittances than those in region 'A'. However, in region 'B' most of the solutions have bunch lengths larger than 1.2 mm, and in region 'C' solutions have kinetic energies away from 5 MeV. Three typical results from these three regions are listed in Table 2. The evident difference among these three parameter settings is the phase of the first RF cavity. Compared to the 'Region A' parameters in Table 2, a smaller RF phase results in lower emittance but with a price of lower beam energy; a larger RF phase leads to an increase in the final emittance and bunch length. It appears that the objectives are in conflict in presence of strong space charge effect. As a result, one should choose a tradeoff between different objectives. Furthermore, the beam distribution in phase space should be optimized to avoid folding structure in $z$ dimension and to make the transverse density profile as close to Gaussian as possible. As a compromise, the chosen result is $E_k$ =5.04 MeV, $\varepsilon_{n,x(y)}$ = 2.35 mm.mrad, $\sigma_z$ = 1.16 mm and $\sigma_\delta$ = 0.56%, with the parameters listed in Table 2 as 'Optimal-0.75' and with the final beam distribution shown in Fig. 8.

Due to the fact that different $\sigma_{laser}$ results in different thermal emittance and different electron density (and different space charge effect), it is necessary to investigate the dependency between the available minimum emittance and the laser beam size. Thus, genetic optimizations for the cases with $\sigma_{laser}$ from 0.3 mm to 1.5 mm are performed. In each case we select the optimal solution that predicts the minimum emittance among those satisfying the condition (3) and resulting in promising distribution in phase space. The variation of the available minimum emittance and the thermal emittance with $\sigma_{laser}$ is presented in Fig. 9. It appears that using an incident laser with $\sigma_{laser}$ of 1 ~ 1.2 mm, it is feasible to achieve an electron beam with emittance below 1 mm.mrad for the high-charge operation mode. During optimization we find that in the cases with too small a laser beam size (e.g. < 0.5 mm), due to high electron intensity and very strong space charge effect, all the solutions in the Pareto front predict relatively large emittance and folding structure in $z$ dimension. On the other hand, too large a laser beam size (e.g. > 1.5 mm) implies a relatively large thermal emittance, which sets the limit of the available minimum emittance. This will cancel out the benefits provided by the low beam intensity and weak space charge effect. In addition, the active area on cathode should be off-axis to avoid the damage due to ion back-bombardment [13]. A larger initial laser beam size requires a larger active area with a larger offset, which will also lead to a greater emittance growth (This has been demonstrated in the beam dynamics study for the low-charge operation mode in Ref. [5]). Also note that the available minimum emittance increases quickly as $\sigma_{laser}$ decreases from 1 mm. Based on the above considerations, an incident laser with $\sigma_{laser}$ of slightly above 1 mm (e.g., 1.1 mm) seems to be the best choice for the high-charge operation mode. Nevertheless, the optimal parameter settings in the case of $\sigma_{laser}$ = 1.0 mm are listed in Table 2 as 'Optimal-1.0', and the final beam distribution is shown in Fig. 10.

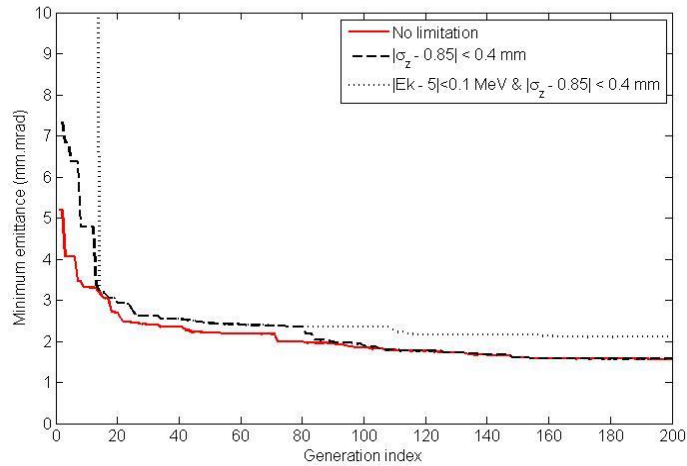

Fig. 6. Evolution the minimum emittances under three different conditions for the high-charge operation mode with $\sigma_{laser}$ of 0.75 mm.

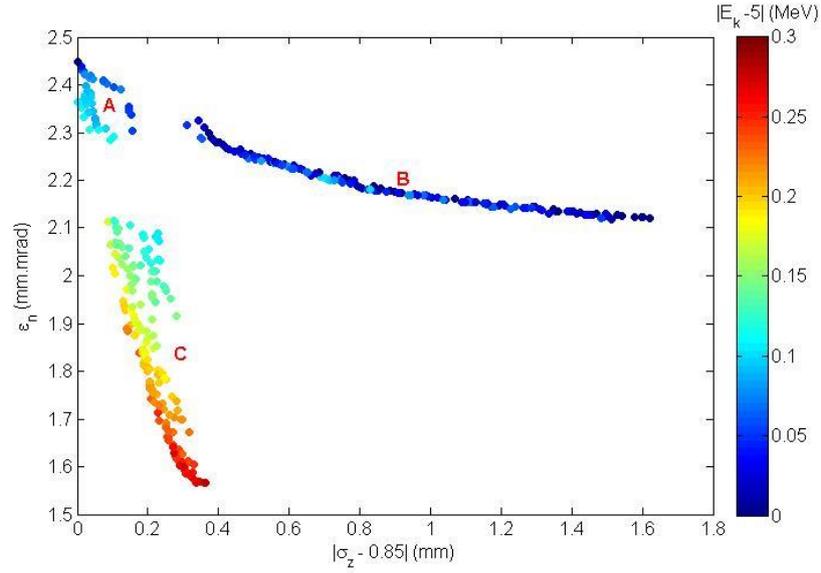

Fig. 7 (color online) Results of the 200th generation in objective space for the high-charge operation mode with $\sigma_{laser}$ of 0.75 mm.

Table 2. Three representative results for high-charge operation mode

| Result | Region A | Region B | Region C | Optimal-0.75 | Optimal-1.0 |
|---|---|---|---|---|---|
| Laser rms tran. size (mm) | | 0.75 | | | 1.0 |
| Final tran. Emittance (mm.mrad) | 2.46 | 2.22 | 1.57 | 2.35 | 0.75 |
| Final rms tran. size (mm) | 2.31 | 1.37 | 5.76 | 1.54 | 1.60 |
| Final rms bunch length (mm) | 0.85 | 1.50 | 0.61 | 1.16 | 1.10 |
| Final beam kinetic energy (MeV) | 5.01 | 5.03 | 4.80 | 5.04 | 4.98 |
| Final rms energy spread (%) | 0.54 | 0.77 | 0.21 | 0.56 | 0.45 |
| 1st solenoid position (m) | 0.24 | 0.24 | 0.24 | 0.24 | 0.24 |
| 1st solenoid peak field (Gauss) | 372.8 | 361.4 | 390.5 | 362.0 | 355.0 |
| Buncher position (m) | 0.816 | 0.816 | 0.803 | 0.814 | 0.80 |
| Buncher peak field (MV/m) | 4.96 | 4.91 | 5.04 | 4.95 | 4.61 |
| Buncher phase (degree) | -160.0 | -160.0 | -160.0 | -160.0 | -138.0 |
| 2nd solenoid position (m) | 1.23 | 1.25 | 1.22 | 1.25 | 1.17 |
| 2nd solenoid peak field (Gauss) | 719.7 | 729.7 | 722.2 | 729.4 | 354.0 |
| 1st cavity position (m) | 1.80 | 1.82 | 1.78 | 1.82 | 1.79 |
| 1st cavity peak field (MV/m) | 19.4 | 19.4 | 19.4 | 17.4 | 67.4 |
| 1st cavity phase (degree) | 6.55 | 16.3 | 0.92 | 19.4 | 21.3 |
| 2nd cavity position (m) | 2.65 | 2.67 | 2.63 | 2.67 | 2.64 |
| 2nd cavity peak field (MV/m) | 20.7 | 20.7 | 20.7 | 20.7 | 19.5 |
| 2nd cavity phase (degree) | 122.7 | 122.9 | 123.0 | 122.8 | 131.0 |

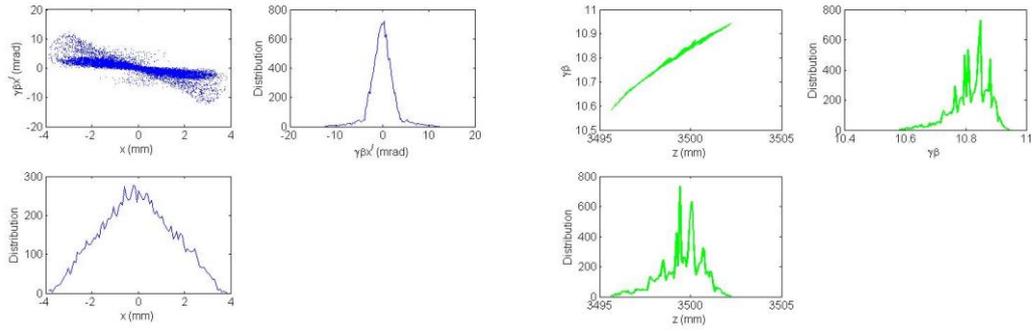

Fig.8. Beam distribution in the phase space of (*x*, *x'*) and (*z*, $E_k$) at the end of the injector with the 'Optimal-0.75' parameters in Table 2.

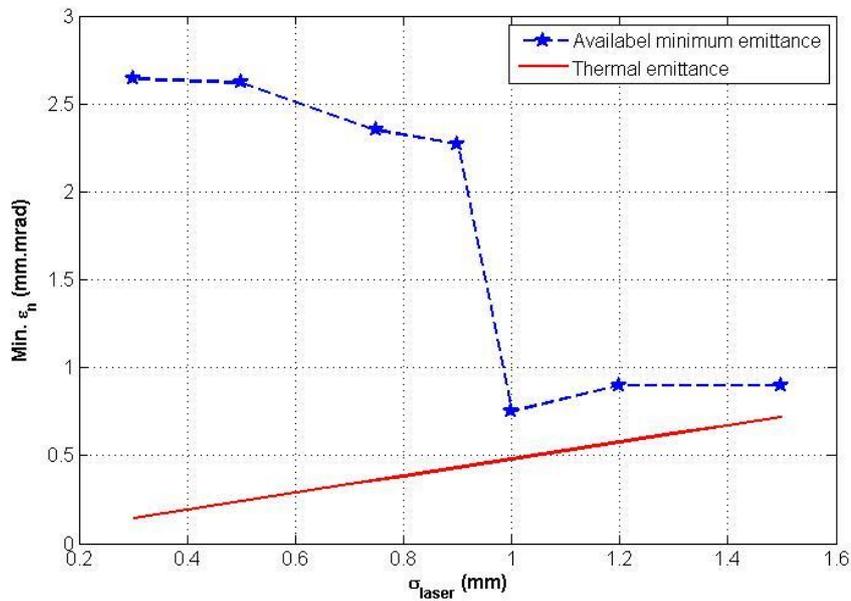

Fig. 9. Variation of the available minimum emittance and the thermal emittance with the laser beam size $\sigma_{laser}$ for the high-charge operation mode.

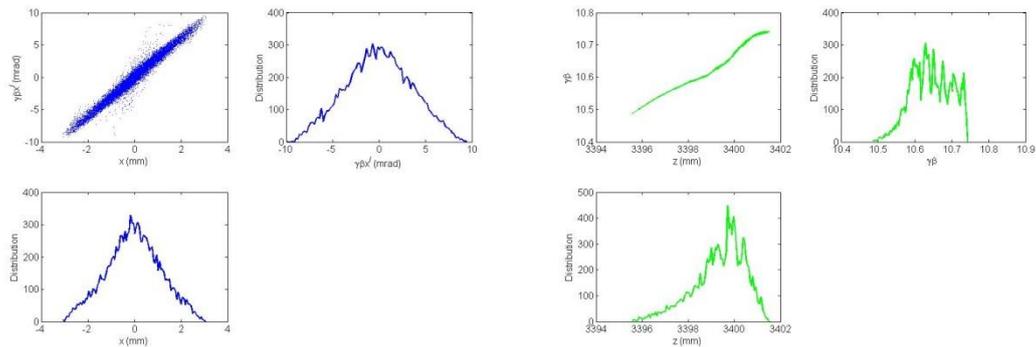

Fig. 10. Beam distribution in the phase space of (*x*, *x'*) and (*z*, $E_k$) at the end of the injector with the 'Optimal-1.0' parameters in Table 2.

**4 Conclusions**

    Based on the beam dynamics study for the ERL-TF injector in low-charge operation mode

which is presented in Ref. [5], in this paper an evolutionary genetic method, non-dominated sorting genetic algorithm II, is applied to optimize the injector beam dynamics, especially in the high-charge operation mode. It appears feasible to achieve an electron beam with kinetic energy of 5 MeV, bunch length of 2 ~ 4 ps, and emittance below 1 mm.mard, at the end of the injector by using an incident laser with rms transverse size of 1 ~ 1.2 mm. It is also found that if releasing the beam energy limitation in some degree, it is possible to obtain relatively small emittance with other laser beam sizes. Above all, these studies will benefit the future construction and commissioning of the ERL-TF injector.

**Acknowledgements**

The author thanks the colleagues of the ERL-TF work group at IHEP for helpful discussions. Thanks also go to LIU Sheng-Guang for sharing the original field data of the injector elements. The colleagues in the Accelerator Physics Group of IHEP are much appreciated for providing computing time for this time-consuming genetic optimization.

**References**

[1] CHEN Sen-Yu, WANG Shu-Hong, ZHU Xiong-Wei. Chinese Physics C, 2010, 34: 112-114

[2] LIU Sheng-Guang, XU Jin-Qiang. Chinese Physics C, 2011, 25(1): 88-91

[3] WANG Shu-Hong, et al. Chinese Physics C, 2012, 36(5): 469-474

[4] CUI Xiao-Hao, JIAO Yi, WANG Jiu-Qing, WANG Shu-Hong, Chinese Physics C, 2013, 37(7): 077005.

[5] JIAO Yi, XIAO Ou-Zheng, Beam dynamics studies of the photo-injector in low-charge operation mode for the ERL test facility at IHEP, Chinese Physics C, to be published.

[6] Qiang J, Lidia S, Ryne R D, Limborg-Deprey C, Phys. Rev. ST Accel. Beams, 2006, **9**: 044204

[7] Bazarov I V, Sinclair C K, Phys. Rev. ST Accel. Beams, 2005, **8**: 034202

[8] Yang L, Robin D, Sannibale F, Steier C, Wan W, Nucl. Instrum. Methods Phys. Res., Sect. A, 2009, **609**: 50-57

[9] Yang L, Li Y, Guo W, Krinsky S, Phys. Rev. ST Accel. Beams, 2011, **14**: 054001

[10] GAO Weiwei, WANG Lin, LI Weimin, Phys. Rev. ST Accel. Beams, 2011, **14**: 094001

[11] Deb K et al. IEEE Transactions on Evolutionary Computation, 2002, **6**(2): 182-197

[12] Bazarov I V, et al. J. Appl. Phys, 2008, **103**: 054901

[13] Dunham B, et al. Appl. Phys. Lett. 2013, **102**: 034105